\documentstyle[12pt,epsfig,wrapfig]{article}
\def\vbs{{\bf b}}
\def\als{\alpha_s}
\newcommand{\LQCD}{\Lambda_{\rm{QCD}}}
\begin{document}
\begin{center}
{\bfseries Leptoproduction of Polarized Vector Mesons}
\vskip 5mm
B.\ Postler
\vskip 5mm
{\small
{\it
Fachbereich Physik, Universit\"at Wuppertal, D-42097 Wuppertal,
Germany
}
}
\end{center}
\vskip 5mm

\begin{abstract}
   We present a status report on a study of vector meson
   leptoproduction for the HERA energy range on the basis of the
   generalized handbag approach for which the amplitudes factorize in
   the parton subprocess $\gamma^* g \to V\, g$ and generalized parton
   distributions (full report in \cite{gkp}). In contrast to the
   leading twist approach transverse degrees of freedom as well as
   Sudakov suppressions are taken into account in the
   subprocess. First results for the cross section of the reaction
   $\gamma_L^*\, p\to\,\rho^0_L\, p$ are found to be in fair agreement
   with experiment.
\end{abstract}

\vskip 10mm

We are going to investigate vector meson leptoproduction off protons at high
energies, large virtuality, $Q^2$, of the exchanged photon and small
invariant momentum transfer, $t$, from the initial to the final
proton. In the kinematical
region we are interested in, the process factorizes into a hard
subprocess - meson leptoproduction off partons - and a soft proton
matrix element which represents a generalized or skewed parton
distribution (GPD) \cite{rad96,col96}. It can be shown that
the cross section is dominated by longitudinally polarized photons for
$Q^2\to \infty$; the cross section for transversely polarized photons
is suppressed by $1/Q$.

The only application of the GPD approach for the reaction has been
performed by Mankiewicz {\it et al.} \cite{mpw}.  It turns out however
that the cross section comes out to large by order of magnitude. This
is similar to the two--gluon exchange approach of Brodsky {\it et al.}
\cite{brodsky} which can be understood as a LO
$\als\rm{log}(Q^2/LQCD)$ calculation, as shown by Frankfurt et
al. \cite{fks}.  The latter authors argue that transverse momentum
effects in the photon wave function lead to a suppression factor
multiplying the leading twist amplitude which leads then to agreement
with experiment.  Motivated by experience with meson form factors
\cite{li92,jak93} we include transverse degrees of freedom as well as
Sudakov suppressions in the subprocess. By these effects contributions
from the end-point regions, in which one of the partons entering the
meson wave function becomes soft and where factorization breaks down,
are suppressed. A similar idea has been advocated by Vanderhaeghen
{\it et al.} \cite{vgg} for the quark subprocess, however both
approaches differ in detail.

\begin{figure}[hbt]
\begin{center}
\psfig{file=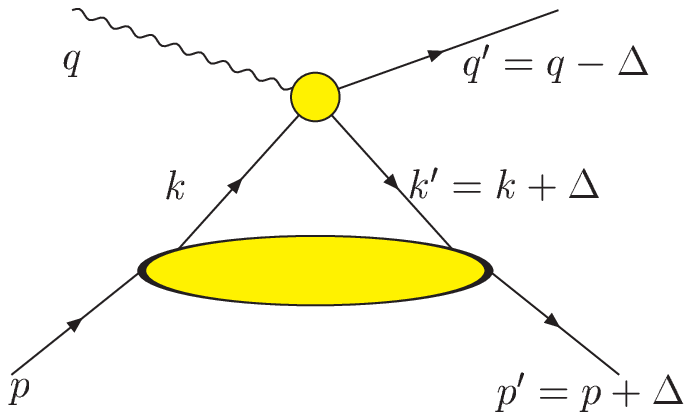, bb=155 550 356 670, height=4cm}
\end{center}
\caption{The handbag-type diagram for leptoproduction of
mesons. The large blob represents a sum over all spectator configuration.
$k$ and $k'$ denote the momenta of the active partons. The small blob
stands for meson leptoproduction off partons.}
\label{fig:1}
\end{figure}

The kinematics of the process are depicted in figure \ref{fig:1}. The
lower blob represents the soft hadron part, which is
parametrized\footnote{Throughout this paper we will use Ji's notation
for GPDs \cite{ji}.} by a
GPD $H(x,\xi,t)$ in case of unpolarized gluons and $\tilde H(x,\xi,t)$
in case of polarized gluons\footnote{The GPDs $H(x,\xi,t)$ and
$\tilde H(x,\xi,t)$ are universal functions and may also be accessed
e.g. in deeply virtual compton scattering.}, whereas the upper blob
contains the hard scattering amplitude and the soft meson production.

Considering the factorization of the process, we will now discuss the
different components of the amplitude of leptoproduction of polarized
vector mesons. For the GPD we use a double distribution ansatz
\cite{rad99a} where a normal parton distribution is convoluted with a
profile function for which we choose $h(x,y) = 6\, y\,(1-x-y)$.  The
meson production is described by a convolution of the distribution
amplitude, for which we took the asymptotic form, and the hard
scattering part. The hard amplitude was calculated considering only
gluonic contributions since the ZEUS data --- with which we compared
our calculations --- are at high enough energies to neglect quark
contributions. 

Furthermore, according to the modified perturbative approach
\cite{li92,jak93} we retain the quark transverse degrees of freedom
and take into account Sudakov suppression\footnote{It has been shown
in \cite{li92,jak93} that factorization is still valid in the modified
perturbative approach.}. The inclusion of these effects leads to
suppression of the contributions from the soft end--point regions
which results in an overall smaller cross section for the 
$\gamma_L^*\, p\to V_L\, p$ process.

The final expression for the amplitude in this approach is
\begin{eqnarray*}
{\cal M}^{V(g)}_{0+,0+} &=& -\frac{2\,e}{\sqrt{N_c}}\,
                \sqrt{1-\xi^2}\,(1+\xi)\,Q\;
                   \int dx d\tau\, \tau \bar{\tau}\, H^g(x,\xi,t)\\[1em]
      &\times& \int d^{\,2} \vbs\, \hat{\Psi} (\tau, -\vbs)\,
                           \hat{T}_g (\tau,Q,\vbs)\,\als(\mu_R)\,
                           \exp{[-S(\tau,b,Q)]}\,,
\end{eqnarray*}
with
\begin{eqnarray*}
   \hat{\Psi}(\tau,\vbs) &=& 2\pi\, \frac{f_V}{\sqrt{2N_c}}\,
                     6\,\tau\bar\tau\,
               \exp{\left[ -\tau\bar{\tau}\frac{\vbs^2}{4a^2_V}\right]}\,,
\\
S(\tau,b,Q)&=&s(\tau,b,Q)+s(\bar{\tau},b,Q)-\frac{4}{\beta_0}
\ln\frac{\ln\left(\mu_R/\Lambda_{QCD}\right)}
                                {\ln\left(1/b\,\Lambda_{QCD}\right)}\,,
\\
 s(\beta,b,Q) &=& \frac{8}{3\beta_0} \left( \hat{q}
                       \ln{\left(\frac{\hat{q}}{\hat{b}}\right)}
                       -\hat{q} + \hat{b} \right) +
    {\rm NLL-terms\quad \cite{DaJaKro:95}}\,,\\
&&     \hat{q} = \ln{\left(\beta Q/(\sqrt{2}\LQCD)\right)}\,,
\quad  \hat{b}= \ln{\left(1/(b\LQCD)\right)}
\end{eqnarray*}
and the hard scattering amplitude is denoted by
$\hat{T}_g (\tau,Q,\vbs)$.

\begin{figure}[hbt]
\begin{center}
\psfig{file=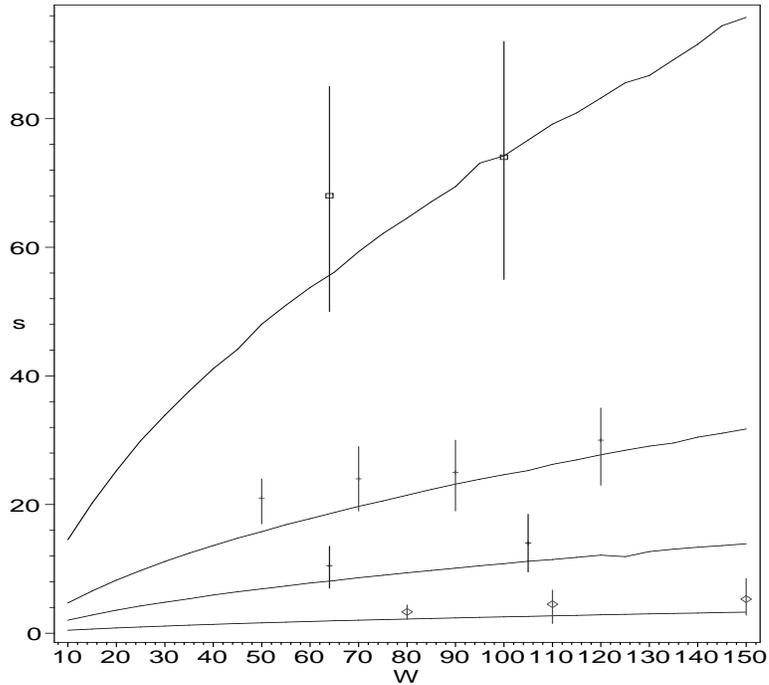, height=9cm,width=10cm}
\end{center}
\caption{The cross section $s(\gamma_L^*\, p\to\,\rho^0_L\, p)$ in nbarn
as a function of the energy $W$ in GeV. Plotted are ZEUS \cite{zeusdata}
datapoints
at $Q^2 = {9,13,17,27}$ GeV$^2$ where the topmost curve has the lowest $Q^2$.
The calculated cross sections are in good agreement with experimental data.}
\label{fig:2}
\end{figure}

As can be seen in figure 2, our calculation describes experimental
data quite well. Without the inclusion of transverse quark momenta and
sudakov suppression the curves would be approximately an order of
magnitude above the data. The modified perturbative approach therefore
is needed to correctly reproduce the data.

Due to the uncertainties, the cross section of vector meson
photoproduction ist not well suited to obtain informations on the
GPDs, in this context the measurement of asymmetries seems more
promising. Especially the measurement of the double--spin asymmetry in
vector meson production --- a task, for which newer experiments like
HERMES are well suited --- may provide interesting insights concerning
the polarized gluon GPDs.

In the future it is planned to calculate all $\gamma^*_\perp \to
V_\perp$ amplitudes --- $\gamma^*_\perp\to V_L$ is already available ---
to examine spin--dependent effects by e.g. calculating the spin density
matrix elements of the $\rho$--meson. In this context it is worthwhile to
note that the mechanism which we propose for the suppression of the
cross section also serves to regularize the infrared divergences, which
appear in the calculation of the $\gamma_\perp^* \to V_L$ amplitudes.

The interest in the $\gamma^*_\perp\to V_L$ transitions bases not only
in possible measurements of spin-dependent observables, e.g. at
COMPASS, but some information on these amplitudes is already available
from the H1 \cite{h1} and ZEUS \cite{zeus} measurements of the decay
density matrix elements in leptoproduction of $\rho$-mesons. The
analysis of these matrix elements demonstrates that both the
$\gamma^*_\perp\to V_\perp$ and the $\gamma^*_\perp\to V_L$ transition
amplitudes are non-negligible even at photon virtualities as large as
$10 \mbox{GeV}^2$. It is common to analyse the density matrix elements
under the assumption of small relative phases between the amplitudes
as is characteristic of Pomeron-type models \cite{iva98}. In this case
the $\gamma^*_\perp\to V_L$ transition amplitudes amount to about
$20\%$ of the dominant $\gamma^*_L\to V_L$ amplitude \cite{h1,zeus}
and, hence, s-channel helicity conservation seems to hold with a
fairly high precision.

Also of interest is the application of our method to the calculation
of the $\Phi$ and $J/\Psi$ photoproduction cross sections which we
intend to do in the near future.


\end{document}